# Link Enhancer for Vehicular Wireless ATM Communications

Arun Kumar S P, Diganta Baishya, Amrendra Kumar
Center for Artificial Intelligence and Robotics
Bangalore, India

{arunkumarsp,diganta_baishya,amrendra}@fastmail.fm

*Majority of the applications used in defense are voice, video and data oriented and has strict QoS requirements. One of the technologies that enabled this is Asynchronous Transfer Mode (ATM) networking. Traditional ATM networks are wired networks. But Tactical networks are meant to be mobile and this implies the use of radio relays for Vehicle-to-Infrastructure (V2I) and Vehicle-to-Vehicle (V2V) communications. ATM networks assume a physical link layer BER of 10-9 or better because of the availability of reliable media like optical fiber links. But this assumption is no longer valid when ATM switches are connected through radio relay where error rates are in the rage of 10-3. This paper presents the architecture of a Link Enhancer meant to improve the Bit Error Rate of the Wireless links used for V2I and V2V communications from 1 in $10^4$ to 1 in $10^8$*

## 1 Introduction

Majority of the applications used in defense are voice, video and data oriented and has strict QoS requirements. One of the technologies that enabled this is Asynchronous Transfer Mode (ATM) networking. Traditional ATM networks are wired networks. But Tactical Networks are meant to be mobile and this necessitates the use of radio relays for V2I and V2V communications. ATM networks assume a physical link layer BER of $10^{-8}$ or better because of the availability of reliable media like optical fiber links. But this assumption is no longer valid when ATM switches are connected through radio relay where error rates are in the rage of $10^{-3}$.

This paper presents the architecture of a Link Enhancer meant to improve the Bit Error Rate of the Wireless links used for V2I and V2V communications from 1 in $10^4$ to 1 in $10^8$. We also discuss a block synchronization mechanism used for synchronization of the blocks for the purpose of successful error control decoding. Our design ensures a probability of false alarm 1 in 10 days, probability of detection 0.9999 and consumes very low (<2%) bandwidth overhead.

Link enhancer allows the usage of commercial ATM switches in V2I and V2V wireless environments. It allows the switches the work efficiently in environments where the Bit Error Rates are of the order of 1 in 1000. Link Enhancer incorporates a Forward Error Correction mechanism and a robust synchronization scheme. A BER improvement from $10^{-3}$ to $10^{-8}$ is provided. The Link Enhancer works transparent to the existing network and therefore can be seamlessly integrated to the network. The system is designed based on System on programmable Chip paradigm wherein the complete logic for cell level processing is implemented in a single Field Programmable Gate Array (FPGA).

Rest of the paper is organized as follows: Section 2 discusses the related work in this area; section 3 describes the system architecture and design details. Section 4 gives the implementation details. Section 5 gives conclusions

## 2 ATM and Wireless Environment

ATM combines the high-speed switching of voice-oriented circuit-switched systems and the flexibility of packet-switched systems in carrying different types of bit-streams – data, voice and video. An ATM network transfers fixed length cells at high speed across point-to-point links. It should be noted that the standard requires that these links are highly reliable with very low bit-error rates (in the order of $10^{-8}$).

But Tactical Networks are meant to be mobile and this implies the use of less reliable media like radio relays for V2I and V2V communications. The motivation for Link Enhancer is to provide appropriate procedure at the wired/wireless network interface to ensure seamless internetworking. A typical tactical scenario using ATM wireless links is shown in Figure 1.

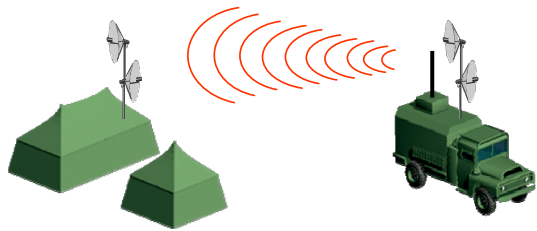

Figure 1(a): V2I Wireless ATM connection

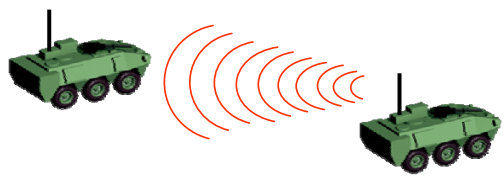

Figure 1(b): V2V Wireless ATM connection

The Link Enhancer works transparent to the existing network and therefore can be seamlessly integrated to the network. The Link Enhancer accepts the E1 data as per ITU-T G.703 [1] & G.704 [2] standards on the input side. The interface to the radio is via G.703 E1 data @ 2.048 Mbps. Figure 2 depicts the usage scenario of Link Enhancer.

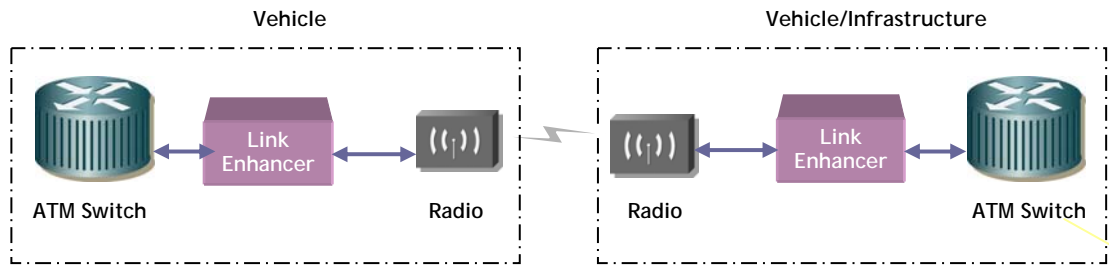

Figure 2: Usage Scenario

## 3 Architecture

The design was captured using Altera Quartus II development software. The tool provided a complete design environment for system-on-a-programmable-chip (SOPC) design. The design was synthesized to generate a gate-level netlist. This netlist was then mapped to the particular technology used in the selected device. This is performed in such a way that the timing constraints are met. Configuration bit stream is then generated which is used to configure the FPGA.

Figure 3 shows the high-level block diagram of the Link Enhancer. It consists of media interfaces – one each for switch and radio interfaces, Cell Delineation and Idle cell management blocks, Reed-Solomon Encoder/Decoder and logic for detection and synchronization for transmitted blocks.

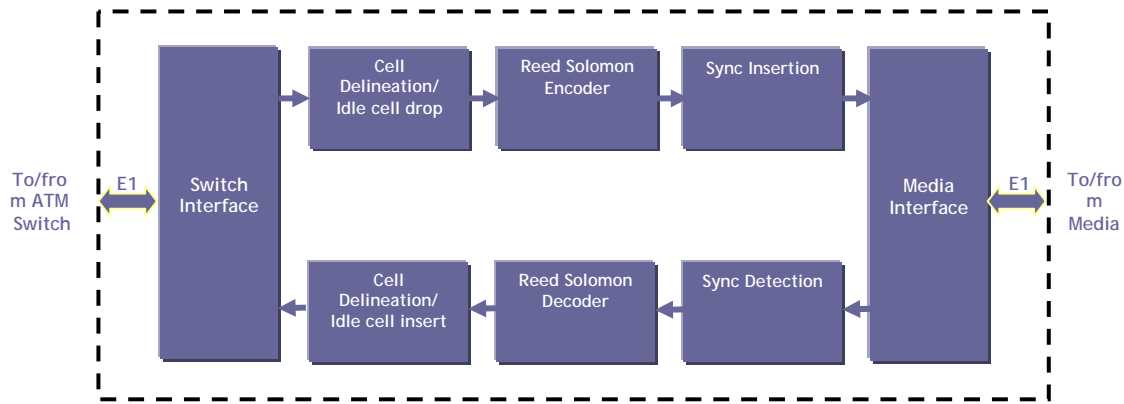

Figure 3: Block Diagram

Figure 4 shows the picture of the completed prototype of the Link Enhancer.

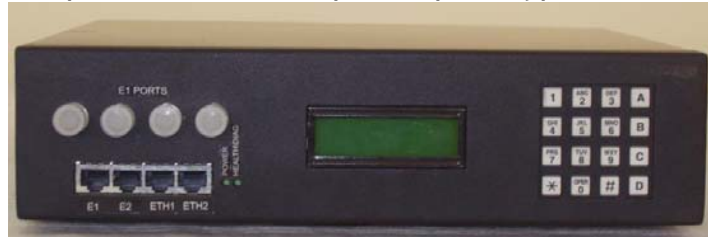

Figure 4: Link Enhancer prototype

### 3.1 Media Interfaces

The Physical Media Dependent functions and E1 frame handling functions are implemented in the interface cards. The interface cards consist of E1 Line Interface Units and Transmit and Receive Framers.

### 3.2 ATM Cell Delineation and Idle Cell Management

Although FEC schemes can be implemented in arbitrary block-level mode, cell-level processing is required in this scenario because of the bandwidth constraints. Due to the extra bandwidth consumed by the encoder redundancy, some data need to be dropped from the incoming stream. Since idle cells do not carry ant valid data, these can be dropped. This calls for an architecture based on cell-level processing. ATM Cell Delineator logic was used to identify the cell boundaries.

The extra bandwidth required for the redundancy added by the encoder is obtained by dropping the idle cells from the incoming stream. The ATM switch generates idle cells when there is no user data. This is to prevent the loss of cell delineation at other end ATM switch. Since valid data is not carried in idle cells, these can be dropped to cater for the required bandwidth at the encoder side. The required rate of idle cells is then inserted after the decoding at the receiver side to prevent loss of continuity, resulting in loss of cell delineation at the switch end.

### 3.3 Reed Solomon Encoder/Decoder

If the reliability provided by the network is lower than the reliability requested by an application, the end systems must make up for the difference. The two basic mechanisms available to improve reliability are *Automatic Repeat reQuest* (ARQ) and *Forward Error Correction* (FEC). ARQ is a *closed-loop* technique based on retransmission of data that were not correctly received by the receiver. ARQ requires the transmitter and receiver to exchange state information about the status of individual messages. Each retransmission of a message adds at least one round-trip time of latency. Therefore, ARQ may not be applicable for transmitting data from applications with low latency constraints. Low latency is necessary for human interaction (voice, video), process control, remote sensing, etc. As the name implies, data for applications using this service is worthless if it does not arrive within a certain time. Another disadvantage of ARQ based schemes is the complexity required to keep track of a potentially very large number of outstanding messages.

Forward Error Correction is an alternative to ARQ that avoids the shortcomings of ARQ and is well suited for operation in high bandwidth-delay product networks. FEC involves the transmission of redundant information along with the original data so that if some of the original data is lost; it can be reconstructed using the redundant information. The amount of redundant information is typically small, so that FEC remains efficient. In data communications, the use of FEC is attractive for providing reliability as needed without increasing the end-to-end latency. FEC can make the operation of the network more cost-effective by allowing it to operate with higher utilization. Without FEC, the network must be operated at a utilization level where the loss rate of the network never exceeds the most stringent loss rate required by any application. In this case, all applications would receive this low loss rate, independent of their actual need.

Because of the above stated reasons, the Forward Error Correction strategy was selected for implementation in Link Enhancer. The study of the performance curves of various error control coding techniques was done and it was observed that Reed-Solomon (RS) codes were the best for the scenario.

RS codes are *non-binary cyclic* codes with symbols made up of *m*-bit sequences, where *m* is any positive integer having a value greater than 2. RS (*n*, *k*) codes on *m*-bit symbols exist for all *n* and *k* for which

$$0 < K < n < 2m + 2$$

where *k* is the number of data symbols being encoded, and *n* is the total number of code symbols in the encoded block. For the most conventional RS (*n*, *k*) code,

$$(n, k) = (2^m - 1, 2^m - 1 - 2t)$$

where *t* is the symbol-error correcting capability of the code, and $n - k = 2t$ is the number of parity symbols. An extended RS code can be made up with $n = 2^m$ or $n = 2^m + 1$, but not any further.

RS codes achieve the *largest possible* code minimum distance for any linear code with the same encoder input and output block lengths. For non-binary codes, the distance between two code-words is defined (analogous to Hamming distance) as the number of symbols in which the sequences differ. For Reed-Solomon codes, the code minimum distance $d_{min}$ is given by:

$$d_{min} = n - k + 1$$

The code is capable of correcting any combination of $t$ or fewer errors, where $t$ can be expressed as:

$$t = \left\lfloor \frac{d_{min} - 1}{2} \right\rfloor = \left\lfloor \frac{n-k}{2} \right\rfloor$$

The above equation illustrates that for the case of RS codes, correcting $t$ symbol errors requires no more than $2t$ parity symbols.

RS codes are particularly useful for *burst-error correction*; that is, they are effective for channels that have memory. Also, they can be used efficiently on channels where the set of input symbols is large. An interesting feature of the RS code is that as many as two information symbols can be added to an RS code of length $n$ without reducing its minimum distance. This extended RS code has length $n + 2$ and the same number of parity check symbols as the original code. The RS decoded symbol-error probability, $P_E$, in terms of the channel symbol-error probability, $p$, can be written as follows

$$P_E \approx \frac{1}{2^m - 1} \sum_{j=t+1}^{2^m - 1} j \binom{2^m - 1}{j} p^j (1-p)^{2^m - 1 - j}$$

where $t$ is the symbol-error correcting capability of the code, and the symbols are made up of $m$ bits each.

The Link Enhancer is designed to improve the Bit Error Rate for $10^{-3}$ to $10^{-8}$ or better. The selected Error Correction Code characterizes this parameter.

The following graph (Figure 5) shows the input error rate to output error rate improvement for various redundancy values. The symbol size is 8 bits restricting the block size to 255 symbols (bytes).

RS is a block code and hence a mechanism must be provided for identification of block boundaries. Synchronization scheme is used for this purpose. A unique synchronization pattern is added before each block before transmitting. The receiver logic can hunt for this pattern and find the block boundaries.

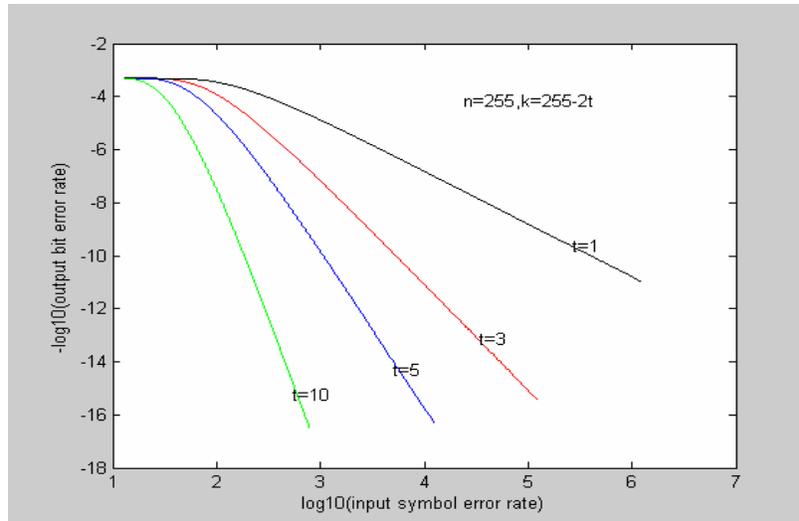

Figure 5: Input error rate vs. Output error rate

## 4 Evaluation and Results

Performance tests were carried out to determine the BER improvement, latency, and work load characteristics. A Network Analyzer was used to generate and monitor ATM E1 traffic. ITU-T O.191 [5] tests were carried out to find various error parameters and the latency. Channel Simulator was used to emulate the radio link characteristics in the laboratory settings. Two setups were used to bring out the performance improvement and latency issues.

4.1 Theoretical Calculations

Cell Error Ratios (CER) corresponding to the required Bit Error Ratios (BER) was also calculated. This is required since the standard tests gave only the CER.

A Cell Error is declared when one or more bits in the ATM payload get corrupted. Payload corruption is detected using CRC in O.191 tests.

$$CER = \sum_{i=1}^{n} \left( {}^{384}C_i \cdot p_e^i \cdot (1 - p_e)^{384-i} \right)$$

n = 384 (ATM Payload size)
$p_e$ = Probability of bit-error ≈ BER
BER = Bit Error Rate
CER = Cell Error Rate

Calculated CER's corresponding to the required BER's are given Table 1:

| BER | CER |
|---|---|
| $10^{-3}$ | 0.31899942 |
| $10^{-8}$ | $3.8399926 \times 10^{-6}$ |

Table 1: BER's and the corresponding CER's

The calculated results are used to compare with the experimental results.

## 4.2 Test Setups

Setup without Link Enhancer, wherein the traffic generated by the network analyzer was routed through the Channel Simulator is shown in Figure 6.

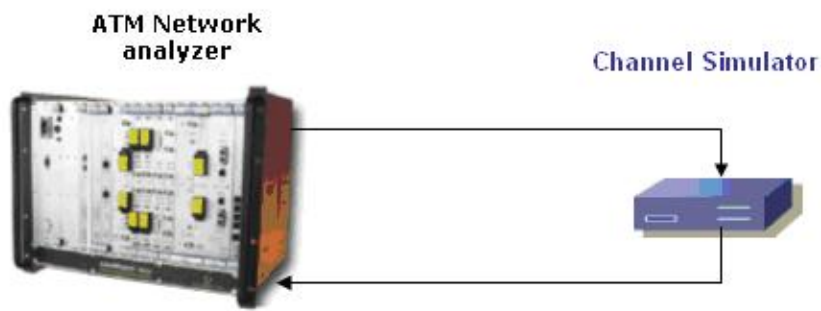

Figure 6: Channel Simulator test setup

Setup with Link Enhancer, where in the Switch Interface was connected to the network analyzer and the Media Interface connected to the Channel Simulator (Figure 7). Full Duplex and Half Duplex testing were performed out.

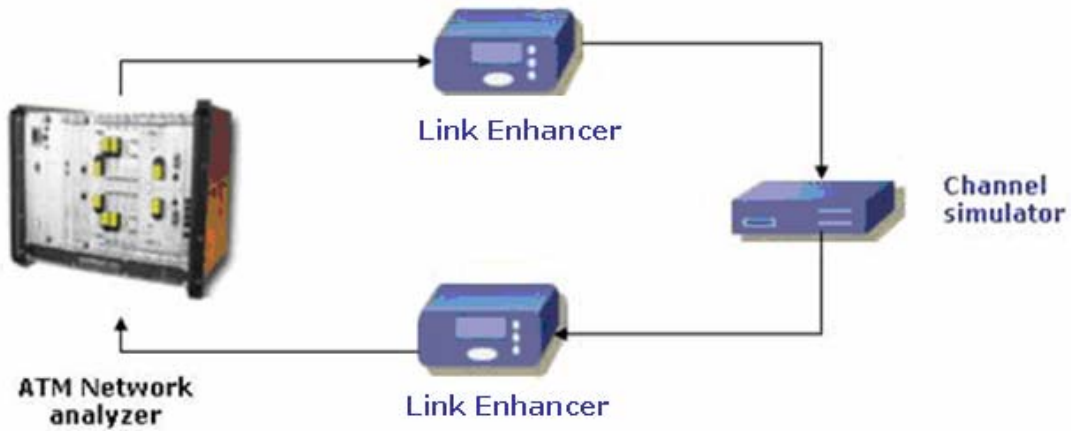

Figure 7: Performance test setup with channel simulator

The following tests were carried out to evaluate the performance of the Link Enhancer:

- BER improvement testing: Measures BER improvements for various input BER's
- Load testing: Measures the system's ability to handle various types of workloads.
- Latency testing: Measures the round trip latency of the transmission path.
- Volume testing: Subjects the system to larger amounts of data to determine its point of failure

4.3 Test Configuration

*Test*: ATM QoS Test (ITU-T O.191)

*Traffic Profile*
- Type: CBR
- Start Idle: 0
- Period: 0
- Utilization: 50
- Bandwidth: 0.959999

*Block Size*
- 16384

ATM QoS Test was run for various time intervals in setup with and without Link Enhancer. The following error-related network performance parameters (defined in Recommendations I.356 [6]) were measured:

- Cell Error Ratio
- Cell Loss Ratio
- Severely Errored Cell Block Ratio

| 1/1000 BER | Without Link Enhancer | With Link Enhancer |
|---|---|---|
| Cells/Sec | 2261.338 | 2264.05 |
| Mbps | 0.9588 | 0.9599 |
| LPAC | OK | OK |
| Total | 2684208 | 2691952 |
| Cells Lost | 1278 | 53 |
| Cell Loss Ratio | 0.000476 | 1.969xE-5 |
| Errored Cells | 847025 | 16 |
| Cell Error Rate | 0.31556 | 5.9436xE-6 |
| Severely Errored Blocks | 163 | 0 |
| Severely Errored Blocks Ratio | 0.75814 | 0 |

Table 2: Performance test results

The Cell Error Ratio observed in all the four experiments and the values obtained by analysis are tabulated in Table 3

| BER | Observed CER | | | | Calculated CER |
|---|---|---|---|---|---|
| | 2mins | 5mins | 10mins | 20mins | |
| $10^{-3}$ | 0.3166 | 0.3161 | 0.3163 | 0.3156 | 0.3189 |
| $10^{-8}$ | $4.08 \times 10^{-6}$ | $9.17 \times 10^{-6}$ | $7.499 \times 10^{-6}$ | $5.94 \times 10^{-6}$ | $3.84 \times 10^{-6}$ |

Table 3: Cell Error Rates: Observed vs. Calculated

The measured values of CER matched with the theoretical calculations.

## 5 Conclusions

RS code (255,235) was found to be best suited for this application because of its efficiency and performance. A suitable block synchronization scheme with very low probability of false alarm and very high probability of detection was designed, implemented and tested. The ATM cell buffer management using the Cell delineation and idle cell detection/insertion modules more or less compensates the overhead introduced by RS code and block synchronization.

The design has been successfully designed, implemented and evaluated for its performance. The results matched with the theoretical analysis.

## 6 References

[1] ITU-T Recommendation G.703 (1991), *Physical/electrical characteristics of hierarchical digital interfaces*.

[2] ITU-T Recommendation G.704 (1995), *Synchronous frame structures used at 1544, 6312, 2048, 8488 and 44 736 kbit/s hierarchical levels*.

[3] ITU-T Recommendation G.804 (1993), *ATM cell mapping into Plesiochronous Digital Hierarchy (PDH)*.